\documentclass[twocolumn,prb]{revtex4}

\usepackage{graphicx,amsfonts,amsbsy,amsthm}
\usepackage{dcolumn}
\usepackage{bm}

\begin{document}
\bibliographystyle{prsty}

\title{Spectroscopic determination of hole density in the ferromagnetic semiconductor Ga$_{1-x}$Mn$_{x}$As}

\author{M. J. Seong$^1$\footnote{Corresponding author: mseong@nrel.gov}, S. H. Chun$^2$,
    Hyeonsik~M. Cheong$^{1,3}$, N. Samarth$^2$, and A. Mascarenhas$^1$}

\address{$^1$ National Renewable Energy Laboratory, 1617 Cole Boulevard, Golden, Colorado 80401\\
$^2$ Department of Physics and Materials Research Institute, The
Pennsylvania State University, University Park, Pennsylvania
16802\\
$^3$ Department of Physics, Sogang University, Seoul
121-742, Korea}

\date{May 23, 2002}

\begin{abstract}
The measurement of the hole density in the ferromagnetic
semiconductor Ga$_{1-x}$Mn$_{x}$As is notoriously difficult using
standard transport techniques due to the dominance of the
anomalous Hall effect. Here, we report the first spectroscopic
measurement of the hole density in four Ga$_{1-x}$Mn$_{x}$As
samples ($x=0, 0.038, 0.061, 0.083$) at room temperature using
Raman scattering intensity analysis of the coupled
plasmon-LO-phonon mode and the unscreened LO phonon. The
unscreened LO phonon frequency linearly decreases as the Mn
concentration increases up to $8.3\%$. The hole density determined
from the Raman scattering shows a monotonic increase with
increasing $x$ for $x\leq0.083$, exhibiting a direct correlation
to the observed $T_c$. The optical technique reported here
provides an unambiguous means of determining the hole density in
this important new class of ``spintronic'' semiconductor
materials.

\end{abstract}

{\parbox{6.5in}{\vspace{-40pt}\textit{Accepted for publication in
Physical Review B (tentatively scheduled to appear in 15 July 2002
issue)}}}

\maketitle


Current interest in the development of a semiconductor
``spintronics'' technology~\cite{wolf01} provides a strong
motivation for fundamental studies of diluted magnetic
semiconductors (DMS)~\cite{furdyna88,awssam99,ohno99}. These are
semiconductors that incorporate magnetic ions such as Mn$^{2+}$
within the crystal lattice. Paramagnetic (and antiferromagnetic)
DMS have traditionally been realized by incorporating isovalent
transition metal ions into II-VI semiconductors such as CdTe and
ZnSe.~\cite{furdyna88,awssam99} The relatively recent discovery of
ferromagnetic III-V semiconductor-based DMS with Curie
temperatures ($T_c$) as high as 110 K has now raised interesting
fundamental issues regarding the origin of ferromagnetism in
materials such as
Ga$_{1-x}$Mn$_{x}$As.~\cite{dietl00,ohno96,deboeck96} In these
III-V DMS, Mn$^{2+}$ acts as an acceptor, generating free holes in
the valence band.~\cite{matsukura98} The ferromagnetism in these
materials arises from the exchange interaction between these holes
and the Mn$^{2+}$ ions, and it is generally believed that there is
a direct correlation between $T_c$ and the hole density
$p$.~\cite{dietl00,konig00} However, the unambiguous determination
of the hole density in Ga$_{1-x}$Mn$_{x}$As by standard
magneto-transport techniques (Hall measurement) is difficult
because of the anomalous Hall effect. The extraction of the
ordinary Hall effect from the measurement, applied at $T<<T_c$,
requires magnetic fields larger than 20 T even at temperatures as
low as 50 mK; even under these conditions, the measured Hall data
are not completely free from the effect of the negative
magnetoresistance, resulting in significant uncertainty in the
deduced hole density. In addition, Hall measurements are not
applicable to magnetically dilute samples that are
insulating.~\cite{ohno99} Finally, we note that the Curie-Weiss
law behavior of the magnetic susceptibility \textit{determined by
the Hall effect} clearly indicates the dominance of the anomalous
Hall effect over the ordinary Hall effect even at room
temperature.~\cite{ohno01} Here, we exploit an alternative method
(Raman scattering) to determine the hole density in
Ga$_{1-x}$Mn$_{x}$As epilayers for a wide range of temperatures by
correlating the hole density to the coupled plasmon-LO-phonon mode
(CPLOM).~\cite{mooradian66,irmer97} Our results show that this
spectroscopic technique provides a reliable method for determining
the hole density in ferromagnetic semiconductors over a broad
range of sample conductivity, ranging from insulating to highly
metallic.

For $n$-type GaAs, the coupling between the LO phonon and electron
plasmon results in two Raman active coupled plasmon-LO-phonon
modes (CPLOM), $L^{+}$ and $L^{-}$.  For high electron density,
$L^{+}$ shows a rapid blue-shift with increasing electron
concentration, providing an accurate calibration for the electron
concentration, whereas $L^{-}$ remains almost stationary near the
GaAs TO frequency. On the other hand, only one CPLOM is observed
in $p$-type GaAs due to a strong hole plasmon damping, moving from
the LO to the TO frequency with increasing hole
concentration.~\cite{irmer97} In this paper, we report the
spectroscopic determination of the carrier concentration of four
Ga$_{1-x}$Mn$_{x}$As samples ($x=0, 0.038, 0.061, 0.083$) at room
temperature using Raman scattering from CPLOM. We find that the
unscreened LO (ULO) phonon frequency of Ga$_{1-x}$Mn$_{x}$As
decreases significantly as the Mn concentration increases up to
$8.3\%$.  This makes the traditional lineshape analysis for a
typical p-type GaAs, where the doping does not change the ULO
frequency, unsuitable for determining the hole density in
Ga$_{1-x}$Mn$_{x}$As. By analyzing the relative Raman intensities
of the ULO phonon and the CPLOM, however, we were able to
determine the carrier concentration up to
$7\times10^{20}~cm^{-3}$. The monotonic increase of the hole
density with increasing $x$ for $x\leq0.083$ correlates well with
the change of $T_c$.


\begin{figure}
       \includegraphics[width=300pt,clip=true,bb=110 195 550 680]{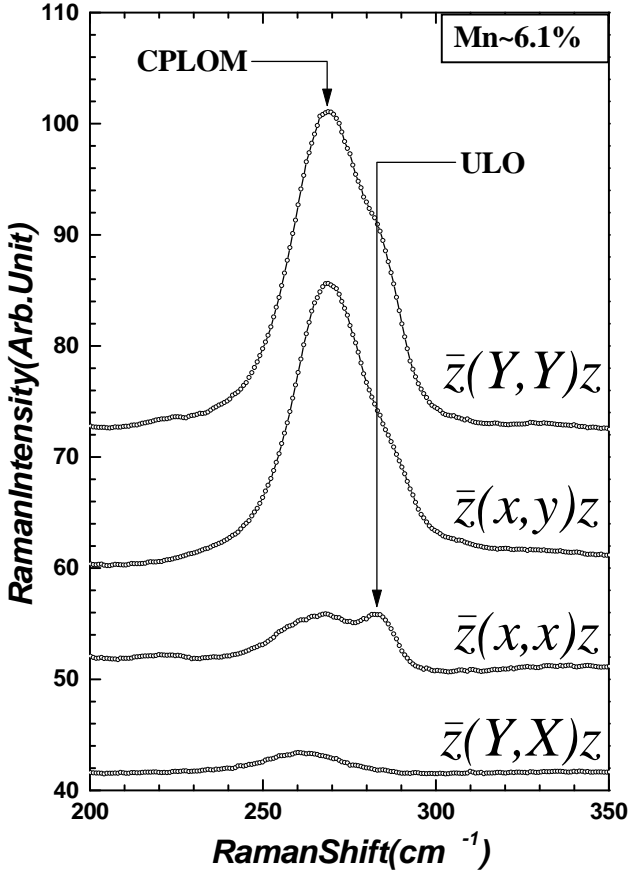}
        \caption{Raman spectra of Ga$_{1-x}$Mn$_{x}$As with 6.1~$\%$ Mn at room temperature for
        $457~nm$ excitation in quasi-backscattering geometry with different polarization configurations,
        where $z=[001]$, $x=[100]$, $Y=[110]$, etc.  The base lines are vertically
        shifted for clarity.
        \label{polarization}}

\end{figure}

Ga$_{1-x}$Mn$_{x}$As epilayers with thickness of $\sim 120~nm$
were grown by molecular beam epitaxy at $\sim 250~^{\circ}C$ on a
(001) semi-insulating GaAs substrate after the deposition of a
buffer structure consisting of a $120~nm$ standard GaAs epilayer
grown at $\sim 550~^{\circ}C$ followed by a $60~nm$
low-temperature grown GaAs epilayer.  Electron microprobe analysis
(EMPA) was used to determine Mn concentrations.  Details about the
growth conditions and parameters are described
elsewhere.~\cite{potashnik01} Raman scattering measurements were
performed at room temperature in a quasi-backscattering geometry
on the (001) growth surface of the samples.  The $457~nm$ line
from a Coherent Ar$^{+}$ laser was used as an excitation light
source in order to obtain a very short penetration depth, thus
avoiding any Raman scattering from the buffer layers. The
scattered photons were dispersed by a SPEX $0.6$-$m$ triple
spectrometer and detected with a liquid-nitrogen-cooled
charge-coupled-device (CCD) detector. The spectrometer was
calibrated using the frequency of the longitudinal optical phonon
peak ($292~cm^{-1}$) of a separate GaAs reference sample.


\begin{figure}
       \includegraphics[width=300pt,clip=true,bb=110 195 550 680]{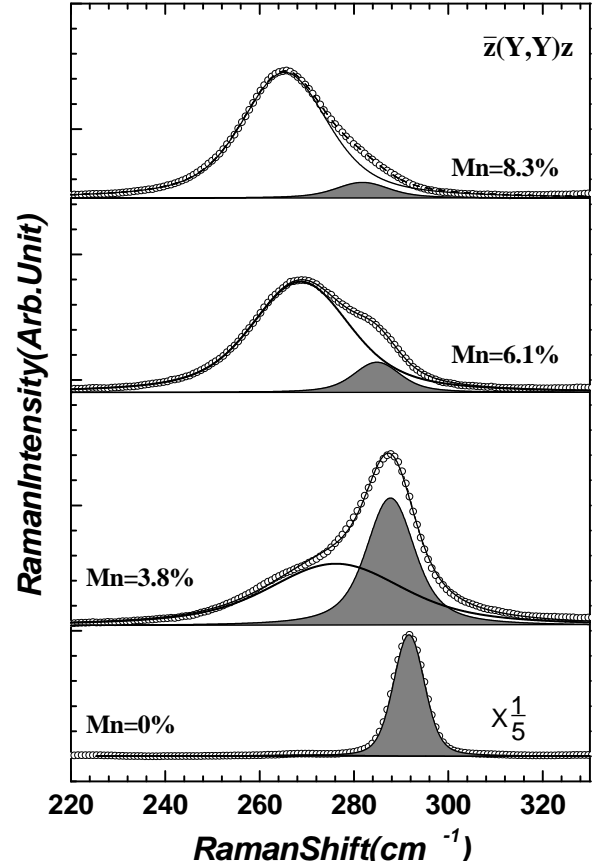}
        \caption{Raman spectra of Ga$_{1-x}$Mn$_{x}$As for $x=0, 0.038, 0.061, 0.083$ at room temperature for
        $457~nm$ excitation in $\overline{z}(Y,Y)z$ scattering configuration, where open circles represent
        experimental data. The shaded area corresponds to the unscreened LO phonon component and the solid
        curves represent contributions from plasmon LO-phonon coupled mode. The base lines are vertically
        shifted for clarity and the Raman intensity of the reference sample was scaled down by 1/5.
        \label{composition}}
\end{figure}

Typical Raman spectra of Ga$_{1-x}$Mn$_{x}$As with 6.1~$\%$ Mn at
room temperature in quasi-backscattering geometry with different
polarization configurations are displayed in
Fig.~\ref{polarization}, where $z=[001]$ is the growth direction
and $x=[100]$, $y=[010]$, $X=[1\overline{1}0]$, and $Y=[110]$.
According to the Raman selection rule for a zinc-blende crystal,
the LO phonon is allowed for $\overline{z}(Y,Y)z$ and
$\overline{z}(x,y)z$ but forbidden for $\overline{z}(x,x)z$ and
$\overline{z}(Y,X)z$ whereas TO phonon is forbidden for all the
scattering configurations employed in
Fig.~\ref{polarization}.~\cite{porto} The Raman feature near
269$cm^{-1}$ is very strong in $\overline{z}(Y,Y)z$ and
$\overline{z}(x,y)z$ whereas it is extremely weak in
$\overline{z}(Y,X)z$ and $\overline{z}(x,x)z$ where LO modes are
forbidden. This reveals its ``LO mode'' nature despite its
proximity to the GaAs TO frequency. The very weak Raman signal
near $\sim 266~cm^{-1}$ in $\overline{z}(Y,X)z$ configuration is
the disorder-induced TO phonon that should exist as a weak
background Raman intensity for all the other scattering
configurations employed in Fig.~\ref{polarization}. Thus, the
Raman feature at 269$cm^{-1}$ is a CPLOM in
Ga$_{0.94}$Mn$_{0.06}$As. We have not observed any Raman signature
in the high frequency spectral range up to 1700~$cm^{-1}$ that can
be attributed to $L^{+}$.  This indicates that the free carrier is
a hole. Apart from the strong CPLOM in $\overline{z}(Y,Y)z$, there
is an unmistakable shoulder on the high frequency side of the
CPLOM. This is due to the ULO in the
depletion layer near the surface. It is more distinctly observed
in $\overline{z}(x,x)z$, where the LO mode is forbidden due to
Raman selection rules. However, the electric field near the
semiconductor surface causes a relaxation of Raman selection
rules.

\begin{figure}
       \includegraphics[width=300pt,clip=true,bb=100 340 550 680]{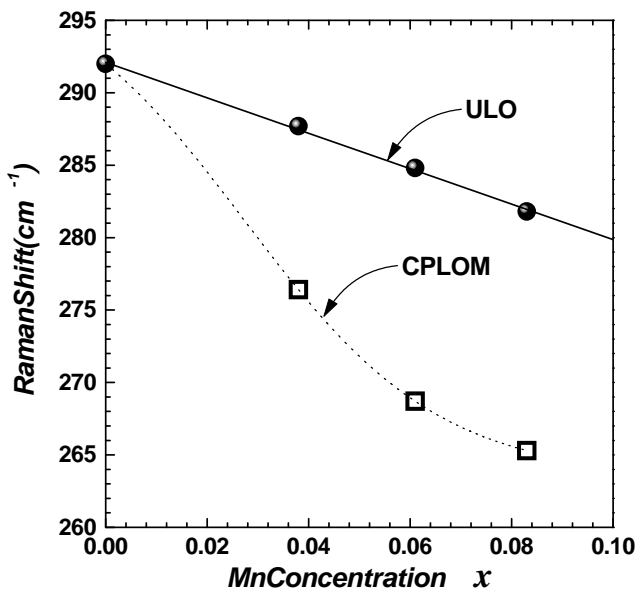}
        \caption{Mn Composition dependence of LO phonon (full circle) Raman frequency
        in Ga$_{1-x}$Mn$_{x}$As ($x\leq0.083$), where the solid line is a linear fit.  CPLOM frequencies are
        also displayed with dashed guide line for eye.
        \label{frequency}}
\end{figure}

In the $\overline{z}(Y,Y)z$ configuration, the superimposed Raman
features can be decomposed into CPLOM and ULO parts by fitting the
the experimental data using two Lorentzian oscillators as shown in
Fig.~\ref{composition}, where Raman spectra of
Ga$_{1-x}$Mn$_{x}$As for $x=0, 0.038, 0.061$, and $0.083$ in
$\overline{z}(Y,Y)z$ scattering configuration are displayed. It
should be noted that the Raman spectrum of the reference sample
($x=0$) consists of only one Lorentzian oscillator. The Raman
intensity of the ULO (shaded area) rapidly decreases as the Mn
concentration increases. The peak positions of the CPLOM and ULO
determined from the curve fitting are listed in
Table~\ref{peakposit} and shown in Fig.~\ref{frequency}. The ULO
frequency linearly decreases with increasing Mn concentrations up
to 8.3\%. Since the lattice constant of Ga$_{1-x}$Mn$_{x}$As
increases with increasing $x$ the compressive strain in the GaMnAs
layer should induce a blue-shift of the ULO frequency. However, the
alloying effect appears to be much stronger than the strain effect
in Ga$_{1-x}$Mn$_{x}$As, leading to the observed ULO frequency
red-shift with increasing $x$.

\begin{table*}[thb]
        \caption{Peak positions of coupled plasmon-LO-phonon mode (CPLOM) and unscreened LO (ULO)
         determined from Raman scattering.  The depletion layer thickness $d$ and the hole concentration $p$
         are calculated using Eqs.~(\ref{d_exp})and (\ref{p}) with $\xi_S \simeq 2$ and $\phi _B = 0.5~V$.
         \label{peakposit}}
        \begin{tabular}{ccccc}
Mn concentration (\%)   &ULO ($cm^{-1}$)    &CPLOM ($cm^{-1}$)
&$d$ (\AA) &$p$ ($cm^{-3}$)\\  \hline
0   &291.7 $\pm$ 1.0   &   &         &  \\
3.8 $\pm$ 0.2   &287.7 $\pm$ 1.0    &276.4 $\pm$ 1.0   &76 $\pm$ 4   &$1.2\pm0.2\times 10^{19}$       \\
6.1 $\pm$ 0.2   &284.8 $\pm$ 1.0    &268.7 $\pm$ 1.0   &16 $\pm$ 1   &2.8$\pm0.4\times 10^{20}$         \\
8.3 $\pm$ 0.2   &281.8 $\pm$ 1.0    &265.3 $\pm$ 1.0   &10 $\pm$
0.5 &7.1$\pm0.7\times 10^{20}$

\end{tabular}

\end{table*}

\begin{figure}
       \includegraphics[width=300pt,clip=true,bb=105 347 590 680]{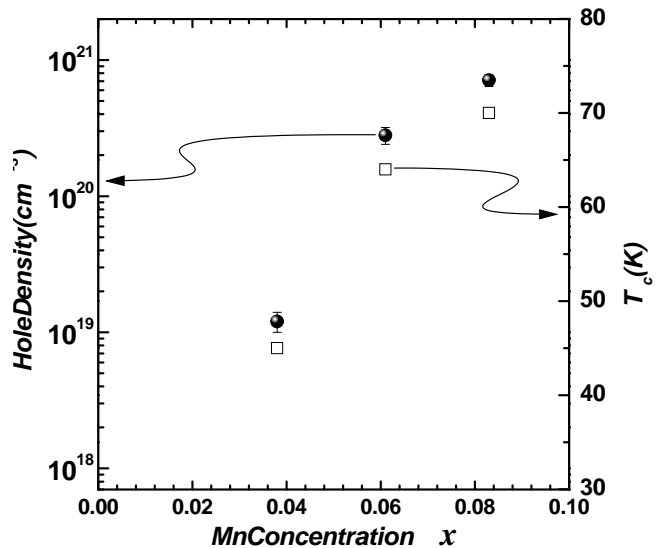}
        \caption{Mn Composition dependence of the hole
        density determined by Raman scattering (full circles) and the ferromagnetic transition
        temperature (open squares) for the same set of Ga$_{1-x}$Mn$_{x}$As samples.
        \label{Tc}}
\end{figure}

Traditionally, lineshape analysis of Raman scattering for the
CPLOM has been used to deduce carrier concentrations of $p$-type
GaAs~\cite{irmer97,mlayah91}, assuming that the phonon frequencies
of the TO and LO phonon do not change with doping. This is valid
because conventional dopant concentrations are too small to change
most of the physical parameters of GaAs used for the lineshape
analysis. However, Mn concentrations in Ga$_{1-x}$Mn$_{x}$As
samples for $p>10^{18}cm^{-3}$ are high enough to change the
frequency of the ULO as shown in Table~\ref{peakposit}, making it
incorrect to use the GaAs parameters for the lineshape analysis of
the CPLOM. Alternatively the $p$-type carrier concentration can be
determined by analyzing the relative intensities of ULO and
CPLOM.~\cite{irmer97} Assuming the Raman scattering efficiency
from the ULO is similar to that in an undoped crystal, the
integrated intensity $A_L$ of the ULO can be written
as~\cite{pinczuk79}
\begin{equation}\label{A_L}
A_L=A_0 \left[ 1-\exp(-2\alpha d) \right],
\end{equation}
where $A_0$ is the integrated intensity in an undoped crystal,
$\alpha$ is the absorption coefficient, and $d$ is the depletion
layer thickness.  Since the integrated Raman intensity is
proportional to the scattering volume, $A_0$ is given by
\begin{equation}\label{A_0}
A_0=\xi_SA_P+A_L,
\end{equation}
where $A_P$ is the integrated intensity of the CPLOM and
$\xi_S=I_L/I_P$ is the relative Raman scattering efficiencies of
the ULO and CPLOM in a unit volume. Using Eqs.~(\ref{A_L}) and
(\ref{A_0}), $d$ can be estimated from the experimental Raman data
\begin{equation}\label{d_exp}
d = \frac{1}{2\alpha}{\mathrm{ln}}(1+\frac{\xi_A}{\xi_S}),
\end{equation}
where $\xi_A=A_L/A_P$ is the ratio of the integrated intensity of
the ULO to that of CPLOM in the Raman spectrum. The depletion
layer thickness $d$ for $p>10^{18}$ can be calculated as a
function of hole concentration $p$ at zero temperature (neglecting
transition region)
\begin{equation}\label{d_theory}
d = \left( \frac{2\varepsilon _0 \varepsilon _S \phi _B}{e}
\right)^{1/2}\frac{1}{p^{1/2}},
\end{equation}
where $\varepsilon _S$ is the static dielectric constant and $\phi
_B$ is the surface potential barrier.~\cite{irmer97} Since the
values of $\varepsilon _S$ and $\phi _B$ for GaMnAs are not
available we used those for GaAs, $\varepsilon
_S=12.8$[Ref.~\cite{blakemore82}] and $\phi _B =
0.5\pm0.05~V$[Ref.~\cite{borisova72}]. By comparing $I_L$ for
$x=0$ and $I_P$ for $x=0.082$ in Fig.~\ref{composition} we have
obtained $\xi_S \simeq 2$ and used this value for the analysis of
all $x$. Since the ULO Raman efficiency in principle could be
dependent on $x$ there is a small uncertainty introduced by using
a constant value of $\xi_S \simeq 2$. But a close inspection
showed that $A_0=(2A_P+A_L)$ is almost constant for all four
samples, making $\xi_S = 2\pm0.1$ a good approximation. We also
used $\alpha \simeq 2.0\times10^5cm^{-1}$ for the excitation
wavelength $457~nm$.~\cite{aspnes83} By equating
Eqs.~(\ref{d_theory}) and (\ref{d_exp}), $p$ is given by
\begin{equation}\label{p}
p =
\frac{8\varepsilon_0\varepsilon_S\alpha^2\phi_B}{e\left[\mathrm{ln}(1+\frac{\xi_A}{\xi_S})\right]^2}~~,
\end{equation}
and thus calculated hole concentrations are listed in
Table~\ref{peakposit}, with an uncertainty less than 10\%. It is
worth mentioning here that any possible uncertainty in $\phi_B$
and $\varepsilon_S$ would affect only the scaling factor in
Eq.~\ref{p}.  In order to check any possible finite temperature
correction in our analysis, we have analyzed the Raman spectrum of
the 6.1\% sample measured at $T=8K$ and obtained the same hole
concentration within the error bar as shown in
Table~\ref{peakposit}. The hole concentration monotonically
increases up to $7\times10^{20}$ for the 8.3\% sample, showing a
good correlation with $T_c$ (Fig.~\ref{Tc}). This is different from the results of
Matsukura {\it et al.}~\cite{matsukura98} where the hole
concentration, measured using Hall effect, and $T_c$ reached its
maximum value $1.5\times 10^{20}~cm^{-3}$ and $110~K$,
respectively, for $x=0.053$ and then decreased with increasing Mn
concentration for $x>0.053$.  The difference between the two
results can be attributed to differences in detailed growth
conditions.  However, the fact that the hole concentration,
determined by Raman scattering, and $T_c$ show a similar monotonic
increase with increasing $x$ provides further confidence in our
spectroscopically determined values of the hole density in
Ga$_{1-x}$Mn$_{x}$As.

In conclusion, we have determined the room temperature carrier
concentration in Ga$_{1-x}$Mn$_{x}$As for $x=0.038,0.061$, and
$0.083$ using Raman intensity analysis of the coupled
plasmon-LO-phonon mode and the unscreened LO phonon.  This study
shows that -- unlike standard Hall measurements -- Raman
scattering provides an unambiguous and reliable method  of
determining the hole density in Ga$_{1-x}$Mn$_{x}$As that can be
profitably exploited for gaining a better understanding of the
origins of ferromagnetism in ferromagnetic semiconductors.


Work at NREL was supported by the Office of Science (Material
Science Division) of the Department of Energy under Contract No.
DE-AC36-99GO10337 as well as the NREL DDRD program. Work at PSU
was supported by DARPA and ONR under grants N00014-99-1-1093,
N00014-99-1-0071 and N00014-99-1-0716. HMC was supported by grant
No. 2000-2-30100-009-3 from the Basic Research Program of the
Korea Science and Engineering Foundation.


\begin{references}
\bibitem{porto}
We have used the conventional Porto's notation $e_i(p_i,p_s)e_s$,
where $e_i$ and $e_s$ are propagation directions of incoming and
scattered photon, respectively, and $p_i$ and $p_s$ are directions
of polarizer and analyzer for incoming and scattered photon,
respectively.
\end{references}
\end{document}